# Nuova frontiera della classificazione testuale: Big data e calcolo distribuito

*Classificazione di token grazie all'ausilio del framework Apache Spark e delle sue relative librerie*

Marco Covelli, Massimiliano Morrelli
Network Contacts - Via Olivetti 17 Molfetta (Ba)

## Abstract

**English**. This document was created in order to study the algorithms for the categorization of phrases and rank them using the facilities provided by the framework Apache Spark. Starting from the study illustrated in the publication "Classifying textual data: shallow, deep and ensemble methods" by Laura Anderlucci, Lucia Guastadisegni, Cinzia Viroli, we wanted to carry out a study on the possible realization of a solution that uses the distributed environment and allows the classification of phrases.

**Italiano.** Il presente documento persegue l'obiettivo di studiare gli algoritmi per la categorizzazione di frasi e classificarle con l'ausilio delle strutture messe a disposizione dal framework Apache Spark. Partendo dallo studio illustrato nella pubblicazione "*Classifying textual data: shallow, deep and ensemble methods*" di Laura Anderlucci, Lucia Guastadisegni e Cinzia Viroli si è voluto realizzare uno studio sulla possibile implementazione di una soluzione in grado di classificare frasi sfruttando i l'ambiente distribuito.

**INDICE**

## INDICE DELLE FIGURE

# 1. Introduzione

Lo studio intrapreso ha come finalità la risoluzione della problematica relativa alla classificazione del contenuto testuale ricavato dalle chiamate ricevute dal servizio clienti di un'azienda operante in ambito *telco* (telefonia mobile) in Italia.
L'analisi è di particolare complessità in quanto in questi ambienti il grado di interazione è talmente elevato, da richiedere una specializzazione nell'analisi dei cosiddetti Big Data e ciò implica l'utilizzo di strumenti consoni per manipolarli e gestirli in modo corretto e puntuale.
L'obiettivo finale del documento è fornire una soluzione più efficiente che garantisca una categorizzazione puntuale ed automatica di determinati token, quando ci si trova di fronte a *dataset* composti da una mole di dati molto elevata, rendendo i tempi di elaborazione sensibilmente inferiori alla norma.

Il documento è stato organizzato seguendo questa struttura:

- la sezione 2 rappresenta una panoramica dei metodi di classificazione convenzionali, le recenti strategie di *deep learning* e i classificatori d'insieme
- la Sezione 3 analizza gli strumenti utilizzati per la risoluzione del problema e le difficoltà riscontrate nell'implementazione del risultato
- la sezione 4 analizza i risultati ottenuti
- la conclusione e le osservazioni finali sono presenti nella sezione 5.

## 2. Stato dell'arte

Secondo la pubblicazione "*Classifying textual data: shallow, deep and ensemble methods*" (1), sono stati designati i migliori algoritmi adibiti alla classificazione dei dati, ed è stato preso in considerazione l'*ensemble*[1] dei suddetti per ottenere una stima di *accuracy* dei risultati.
La prima implementazione degli algoritmi è stata realizzata nel linguaggio di programmazione R che è l'ambiente di sviluppo specifico di riferimento per l'analisi statistica dei dati.
La fase preliminare ha riguardato l'installazione del software *RStudioServer* su un elaboratore così da poter esporre un servizio in grado di addestrare i modelli ed effettuare previsioni attraverso il metodo ensemble e interfacciarsi con il linguaggio Java attraverso chiamate di tipo *REST*.

### 2.1 Fasi di processo

Le fasi di processo sono principalmente due:

- fase di addestramento
- fase di predizione.

### 2.2 Limiti progettuali

Naturalmente, l'introduzione di R come ambiente di riferimento ha portato non poche problematiche in fase di implementazione:
- I dati possono essere analizzati solo se strutturati in modelli specifici (vettori, matrici, ecc..)
- la conversione non è agevole e richiede diversi passaggi
- è obbligatorio installare l'ambiente R per poter utilizzare le sue funzioni in una Java Application
- ogni funzione della libreria *Rserve* genera una nuova connessione su RServer
- richiamare oggetti dinamicamente risulta complesso
- mancanza di automatismi guidati per la configurazione del servizio *Rserve*
- rallentamento causato dal caricamento dei modelli.

### 2.3 Ingegnerizzazione algoritmi di R

Per superare i suddetti vincoli si è deciso quindi di reingegnerizzare tutti gli algoritmi attraverso il linguaggio Java, escludendo l'intervento di RServer, sia per diminuire il tempo di elaborazione, sia per aumentare la potenza di calcolo attraverso l'impiego del sistema distribuito. L'elaborato quindi, invece di essere processato da un singolo server, è stato suddiviso e processato in maniera parallela da diverse macchine ottenendo risultati davvero soddisfacenti.
Apache Spark è il framework utilizzato per amministrare e gestire il calcolo distribuito.

---

[1] Il metodo *ensemble* sarà descritto nel paragrafo 3.6

# 3. Metodi di classificazione

## 3.1 Naive Bayes

La classificazione *Bayesiana* è una tecnica statistica con la quale si determina la probabilità di un elemento di appartenere a una certa classe.

La tecnica si basa sul teorema di Bayes particolarmente adeguata quando la dimensione degli input è elevata. Il teorema definisce la probabilità condizionata (o a posteriori) di un evento rispetto ad un altro.

I classificatori di *Naive Bayes* possono gestire un numero arbitrario di variabili indipendenti, sia continue che discrete. Dato un insieme di variabili, $X = \{x_1, x_2, …, x_d\}$, si vuole costruire la probabilità a posteriori per l'evento $c_j$ tra un insieme di possibili eventi $C = \{c_1, c_2, …, c_d\}$.

In un linguaggio più familiare, X è il predittore e C è l'insieme di livelli categoriali presenti nella variabile dipendente. Utilizzando la regola di Bayes:

$$p(c_j|x_1, x_2, …, x_d) \propto p(x_1, x_2, …, x_d|c_j)p(c_j) \qquad \text{Eq. 2.1}$$

dove $p(c_j|x_1, x_2, …, x_d)$ è la probabilità a posteriori di appartenenza alla classe, cioè la probabilità che $x$ appartenga a $c_j$. Poiché Naive Bayes presuppone che le probabilità condizionali delle variabili indipendenti siano statisticamente indipendenti, possiamo scomporre la probabilità in un prodotto di termini:

$$p(X|c_j) \propto \prod_{k=1}^{d} p(x_k|c_j) \qquad \text{Eq. 2.2}$$

Ed è possibile riscrivere la probabilità a posteriori come:

$$p(c_j|X) \propto p(c_j) \prod_{k=1}^{d} p(x_k|c_j) \qquad \text{Eq. 2.3}$$

Usando la regola di Bayes appena descritta, si riesce a etichettare un nuovo caso $x$ con un evento $c_j$ che raggiunge la massima probabilità a posteriori (1).

## 3.2 K-nearest neighbors

L'algoritmo *k-nearest neighbors* (KNN) è un algoritmo di apprendimento automatico supervisionato utilizzato per risolvere sia problemi di classificazione che di regressione.
Un algoritmo di apprendimento automatico supervisionato si basa su dati di input etichettati per addestrare una funzione che produce un output appropriato quando vengono forniti nuovi dati senza etichetta, ovvero, data una coppia di valori $(x, y)$ sarà necessario trovare una relazione fra di loro. In particolare si dovrà addestrare una funzione $h = X \to Y$ che permetterà di poter conoscere il valore y tramite la funzione $h(x)$.
L'algoritmo KNN presuppone che oggetti simili esistano nelle immediate vicinanze, in altre parole, cose simili sono vicine l'una all'altra. L'algoritmo KNN dipende da questo presupposto, e riesce ad unire il concetto di somiglianza (a volte chiamata distanza, prossimità o vicinanza) e reinterpretarlo in una funzione matematica.

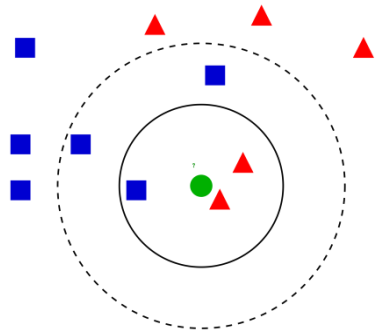

*Figura 3.1 - k-esima suddivisione per la classificazione k-nn di un spazio contenente elementi*

Il classificatore KNN è anche un algoritmo di apprendimento non parametrico e basato su istanze. Il significato di non parametrico sta nella eliminazione delle assunzioni esplicite sulla formula funzionale di h, così facendo si evita il pericolo di modificare la distribuzione sottostante dei dati.
Mentre l'apprendimento basato su istanze implica che l'apprendimento non avviene tramite l'addestramento esplicito di un modello ma si sceglie di memorizzare le istanze[2] di addestramento che costituiscono la "conoscenza" utile alla predizione (2).
L'algoritmo KNN si riduce essenzialmente a valutare la somiglianza degli elementi più vicini e, a maggioranza, classificare l'elemento. La somiglianza viene definita in base a una metrica di distanza tra due punti di dati. Una scelta popolare è la distanza euclidea data da:

$$d(x, x') = \sqrt{(x_1 - x'_1)^2 + (x_2 - x'_2)^2 + \cdots + (x_n - x'_n)^2} \qquad Eq.\ 2.4$$

ma altre è possibile valutare misure più adatte al proprio dataset ed è possibile scegliere tra:

- Manhattan;
- Chebyshev;
- Hamming.

---

[2] Istanza: punto in uno spazio n-dimensionale.

## 3.3 Support Vector Machine

SVM o *Support Vector* Machine è un modello lineare per problemi di classificazione e regressione. Può risolvere problemi lineari e non lineari e funziona bene per molti problemi pratici. L'idea del modello SVM è semplice: l'algoritmo crea una linea o un iperpiano che separa i dati in classi.

In prima approssimazione ciò che SVM fa è trovare una linea di separazione (o iperpiano) tra i dati di due classi. SVM è un algoritmo che prende i dati come input e genera una linea che separa tali classi, se possibile.

Secondo l'algoritmo i punti più vicini alla linea di entrambe le classi, sono chiamati vettori di supporto (3).

Chiamando $w$ il vettore normale all'iperpiano e $b$ l'intercetta all'origine, tutti i punti che giacciono sull'iperpiano separatore dovranno soddisfare l'equazione:

$$w \cdot x + b = 0 \qquad \text{Eq. 2.5}$$

e la funzione di decisione corrispondente sarà:

$$f(x) = \text{sign}(w \cdot x + b) \qquad \text{Eq. 2.6}$$

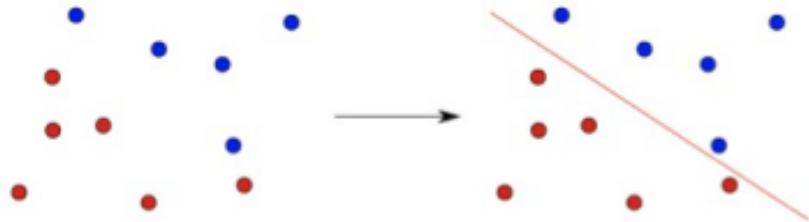

*Figura 3.2 - Classificazione binaria lineare*

Nella maggior parte dei casi pratici bisogna anche tener conto che ci possono essere degli errori nei dati sperimentali. Se ci sono dei punti sbagliati nel semipiano, infatti, una classificazione lineare esatta è pressoché impossibile e quindi sono necessarie tecniche di approssimazione che rendano minimo il numero di errori (4).

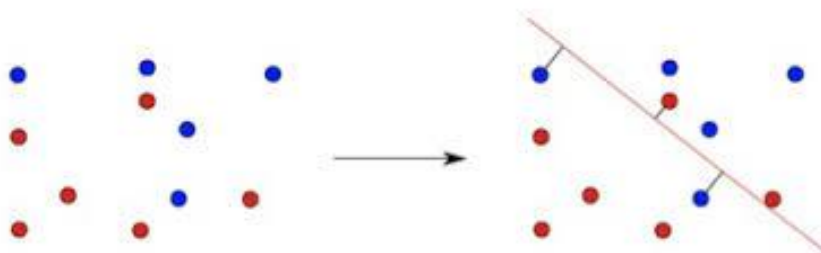

*Figura 3.3 - Classificazione binaria lineare con errori*

Ora calcoliamo la distanza tra la linea e i vettori di supporto. Questa distanza è chiamata margine. L'obiettivo è massimizzare il margine. L'iperpiano per il quale il margine è massimo è l'iperpiano ottimale.

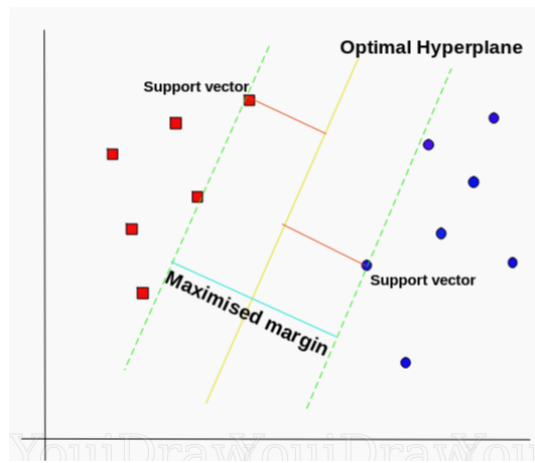

*Figura 3.4 - Iperpiano ottimale utilizzando l'algoritmo SVM*

## 3.4 Random Forest

Gli alberi decisionali sono approcci non parametrici che consentono di rappresentare un certo numero di regole di classificazione riferite a singole variabili prese una alla volta e traslate in una struttura ad albero (5).

La struttura è composta da diversi elementi: i **nodi interni** rappresentano le caratteristiche, ogni ramificazione della struttura rappresenta la migliore divisione ottenibile per una data caratteristica presa come riferimento, secondo un determinato criterio. Le **foglie** rappresentano le classi.

Il numero di rami cresce in modo esponenziale con il numero di funzioni prese in considerazione. Per questo motivo tale metodo può risentire di un numero eccessivo di funzionalità.

Il vantaggio principale dell'albero decisionale è la sua semplicità di lettura e nella sua interpretazione, anche per i non esperti. Il classificatore *Random Forest* si basa sull'idea di costruire una moltitudine di alberi di classificazione nella fase di addestramento (6).

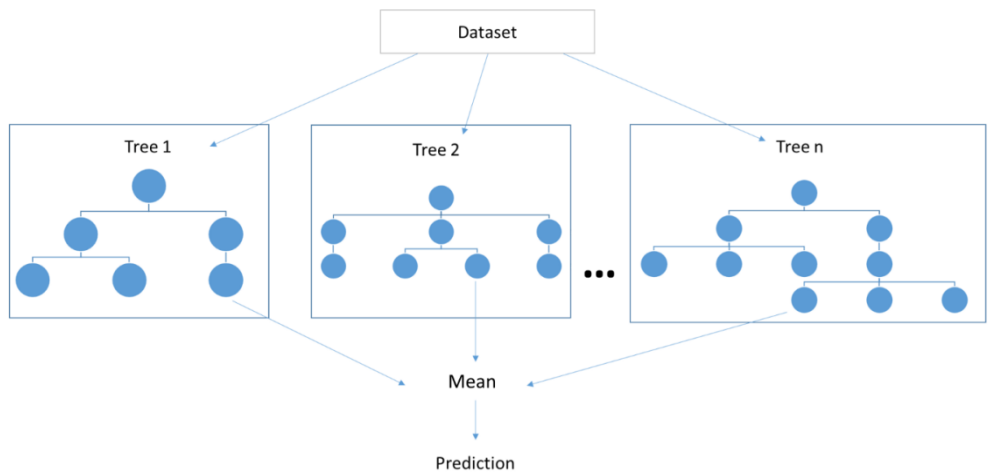

*Figura 3.5 - Albero decisionale Random Forest*

## 3.5 Deep learning

Il *Deep Learning* (noto anche come apprendimento strutturato profondo o apprendimento gerarchico) fa parte di una più ampia famiglia di metodi di Machine Learning basati sull'assimilazione di rappresentazioni di dati, al contrario di algoritmi per l'esecuzione di task specifici.

Con il Deep Learning vengono simulati i processi di apprendimento del cervello biologico attraverso sistemi artificiali (le reti neurali artificiali nello specifico) per insegnare alle macchine non solo ad apprendere autonomamente, ma a farlo in modo più "profondo" paragonabile ad un cervello umano. Naturalmente *profondo* si intende "su più livelli" (vale a dire sul numero di layer nascosti nella rete neurale – chiamati hydden layer).

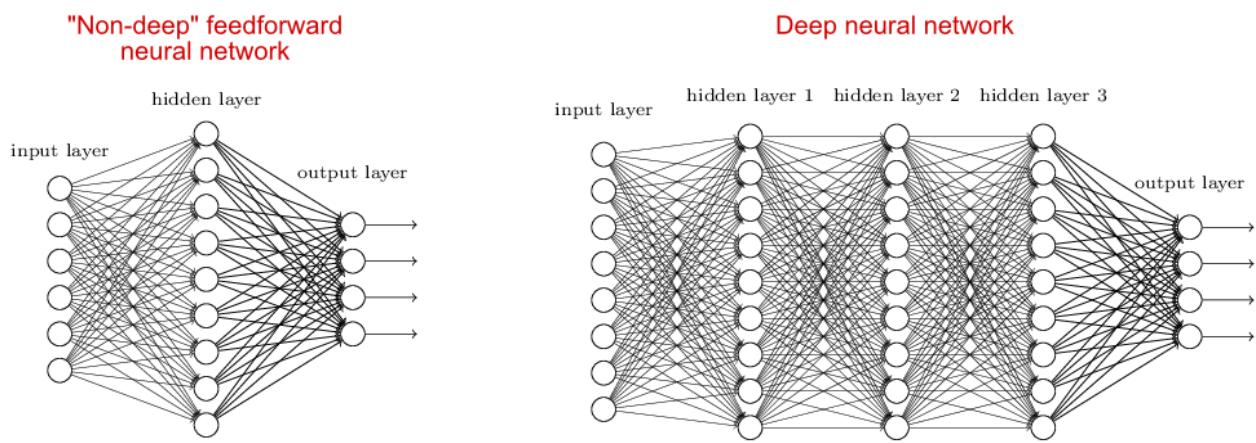

*Figura 3.6 - Differenza fra reti neurali classiche e deep learning*

Le reti neurali profonde sfruttano un numero maggiore di strati intermedi (hydden layer) per costruire più livelli di astrazione, proprio come si fa nei circuiti booleani (modello matematico di computazione usato nello studio della teoria della complessità computazionale che, in informatica, afferisce alla teoria della computabilità ossia studia le risorse minime necessarie per la risoluzione di un problema) (7).

## 3.6 Metodo Ensemble

Il concetto di addestramento Ensemble richiama l'utilizzo di differenti classificatori, che collaborano, attraverso specifiche operazioni, allo scopo di massimizzare le prestazioni utilizzando i punti di forza di ogni singolo modello e limitando le debolezze di ognuno. Alla base del concetto di Ensemble Learning ci sono i classificatori deboli (weak classifier).

Sommati in un certo modo tra di loro, i classificatori deboli permettono di costruire un classificatore forte, risolvendo allo stesso tempo problemi tipici dei classificatori tradizionali (overfitting in primis). Potremmo ad esempio generare cento alberi decisionali differenti partendo dallo stesso insieme di addestramento e farli votare sulla classificazione migliore di un nuovo evento (8) .

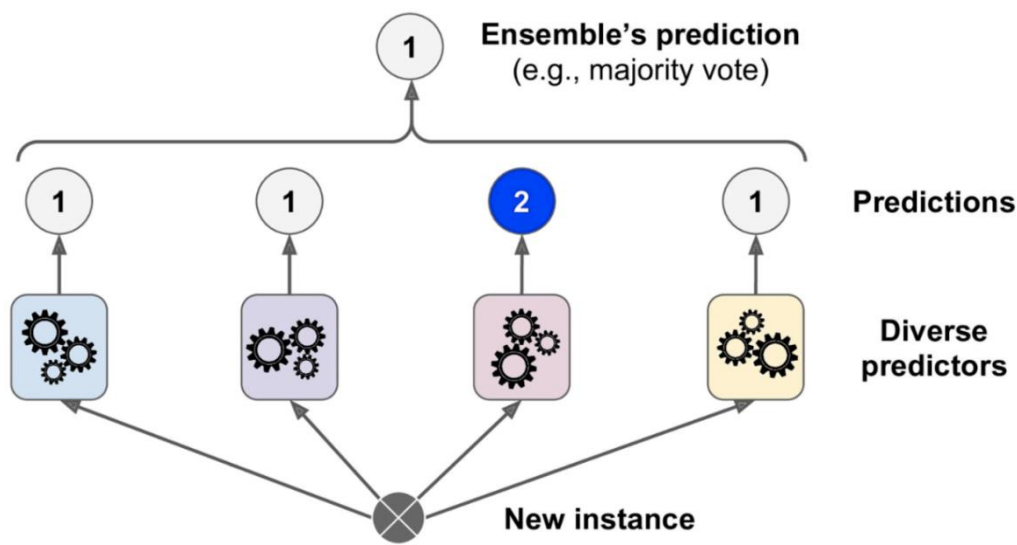

*Figura 3.7 - Ensemble's prediction*

Nel machine learning l'apprendimento Ensemble si divide in tre tecniche fondamentali:

- **Bagging**: Questa tecnica mira a creare un insieme di classificatori aventi la stessa importanza. All'atto della classificazione, ciascun modello voterà circa l'esito della predizione e l'output complessivo sarà la classe che avrà ricevuto il maggior numero di voti.
- **Boosting**: A differenza del bagging, ciascun classificatore influisce sulla votazione finale con un certo peso. Tale peso sarà calcolato in base all'errore di accuratezza che ciascun modello commetterà in fase di learning.
- **Stacking**: Mentre nel bagging l'output era il risultato di una votazione, nello stacking viene introdotto un ulteriore classificatore (detto meta-classificatore) che utilizzerà le predizioni di altri sotto-modelli per effettuare un ulteriore learning.

# 4. Studio delle tecnologie esistenti

Per poter implementare i metodi di classificazione descritti nel capitolo precedente il team di ricerca ha utilizzato la libreria di machine learning messa a disposizione da Apache ovvero Apache MLlib.

La libreria non implementa tutti i predittori citati precedentemente. È stato necessario, quindi, utilizzare la libreria Deeplearning4j per realizzare una rete artificiale che sfruttasse i vantaggi del calcolo distribuito.

## 4.1 Machine Learning Library (MLlib)

Apache Spark è un sistema di calcolo distribuito eseguito da un cluster formato da diversi elaboratori. Fornisce API di alto livello in Java, Scala, Python e R e un motore ottimizzato che supporta grafici di esecuzione generale.

Supporta anche un ricco set di strumenti di livello superiore tra cui:

- **Spark SQL** è utilizzato per l'elaborazione di dati strutturati;
- **MLlib** per l'apprendimento automatico;
- **GraphX** per l'elaborazione di grafici;
- **Spark Streaming** (9).

MLlib è la libreria di machine learning (ML) di Spark. Il suo obiettivo è rendere l'apprendimento automatico pratico facile e scalabile. Ad un livello elevato, fornisce strumenti come:

- **Algoritmi ML:** algoritmi di apprendimento comuni come classificazione, regressione, clustering e filtraggio collaborativo
- **Caratterizzazione:** estrazione delle caratteristiche, trasformazione, riduzione della dimensionalità e selezione
- **Pipelines:** strumenti per la costruzione, la valutazione e l'ottimizzazione delle pipeline ML
- **Persistenza:** salvataggio e caricamento di algoritmi, modelli e pipeline
- **Utilità:** algebra lineare, statistiche, gestione dei dati, etc. (10)

## 4.2 Deeplearning4j

Deeplearning4j è la prima libreria open-source distribuita, scritta per Java e Scala. Integrato con Hadoop e Apache Spark, DL4J porta l'intelligenza artificiale in ambienti aziendali da utilizzare su GPU e CPU distribuite.

A differenza di altri software di apprendimento automatico, DL4J esegue il caricamento dei dati e degli algoritmi di training in processi separati. Rispetto alle diverse API di machine learning, il modello viene caricato in memoria utilizzando i DataVec[3], fornendo maggior flessibilità e conservando la comodità di un semplice upload dei dati.

Prima che l'algoritmo possa iniziare l'apprendimento, è necessario modellare i dati, anche se si dispone già di un modello addestrato. Con modellazione si intende caricare i dati e trasformarli nella giusta forma e nell'intervallo di valori (es. normalizzazione, media zero o varianza dell'unità).

---

[3] DataVec risolve uno degli ostacoli più importanti nel deep learning, permette di ottenere dati un formato che le reti possano comprendere e manipolare, in particolare in formato vettoriale.

## 4.3 Criticità nell'implementazione

### 4.3.1 Naive Bayes, Random Forest, Support Vector Machine

L'implementazione dei metodi non ha evidenziato criticità nel suo sviluppo poiché è bastato seguire le linee guida presenti e descritte nella libreria *MLlib* [4]. La documentazione reperibile sul sito https://spark.apache.org/docs/2.3.1/ml-guide.html, è molto esaustiva. I metodi sono stati implementati seguendo la configurazione presente in (1).

L'unica difficoltà riscontrata in fase di implementazione, è stata l'analisi e la replicazione del metodo Support Vector Machine che, per sua natura, è un classificatore binario. Per poter sfruttare la moltiplicità di classi è stato quindi necessario integrare SVN con *"One-vs-Rest classifier":* metodo della libreria che permette di utilizzare un classificatore binario quando si ha un numero di classi superiore a 2.

### 4.3.2 Deeplearning4j

Per implementare la soluzione è stato definito un modello di building seguendo lo studio effettuato nella ricerca descritta in (1) e successivamente si è utilizzato un approccio di tipo "trial and error" per adattarlo, ottenendo un accuracy maggiore o uguale allo 0.98.

Nello specifico si è configurata la rete neurale come di seguito:

```
MultiLayerConfiguration conf = new NeuralNetConfiguration.Builder()
                                    .seed(Const.DeepProperty.SEED)
                                    .optimizationAlgo(OptimizationAlgorithm.STOCHASTIC_GRADIENT_DESCENT)
                                    .updater(new Adam())
                                    .list()
                                    .layer(0, new DenseLayer.Builder()
                                                .activation(Activation.ELU)
                                                .nIn(numInputs)
                                                .nOut(Const.DeepProperty.UNIT)
                                                .build())
                                    .layer(1, new OutputLayer.Builder(LossFunctions.LossFunction.RECONSTRUCTION_CROSSENTROPY)
                                                .activation(Activation.SOFTMAX)
                                                .nIn(Const.DeepProperty.UNIT)
                                                .nOut(Const.Path_Dataset.CATEGORY.size())
                                                .build())
                                    .build();
```

---

[4] La javadoc della libreria è presente all'indirizzo: "https://spark.apache.org/docs/2.3.1/ml-guide.html"

Per poter sfruttare l'ambiente distribuito per addestrare il modello, è stato necessario creare un oggetto di tipo *TrainingMaster* così definito:

```
TrainingMaster tm = new ParameterAveragingTrainingMaster.Builder(Const.DeepProperty.BATCHSIZE)
                            .averagingFrequency(5)
                            .workerPrefetchNumBatches(Const.DeepProperty.BATCHSIZE)
                            .batchSizePerWorker(Const.DeepProperty.BATCHSIZE)
                            .build();

SparkDl4jMultiLayer sparkNet = new SparkDl4jMultiLayer(jsc, conf, tm);
```

Diversamente da quanto descritto in (1) si è scelto un addestramento di tipo "one-shot" ovvero, sono state definite le epoche in maniera unitaria cioè pari a 1. L'uso di una numerosità di epoche elevata sovraccarica il modello quindi il "rumore" nei dati addestrerà in modo improprio il modello.

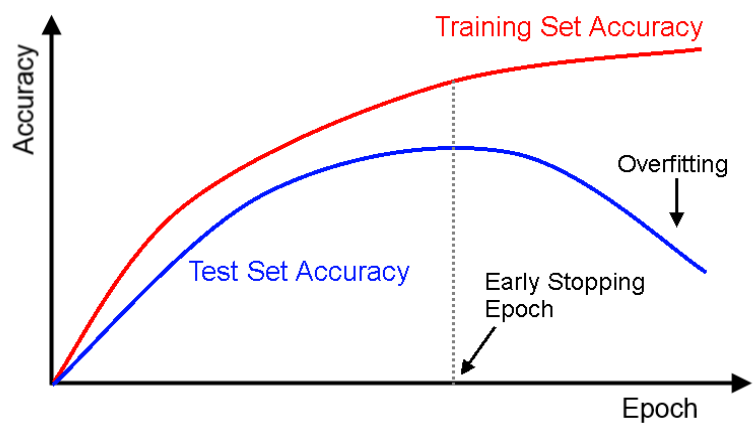

*Figura 4.1 - Andamento dell'addestramento in funzione delle epoche*

### 4.3.3 K-nearest neighbors e Metodo Ensemble

Per poter implementare il metodo KNN non è stato possibile utilizzare librerie ad hoc, ma è stato prodotto in casa, un metodo di classificazione in ambito distribuito.

In particolare si è utilizzato il metodo "*cosineDistanceMisure.distance*" della libreria "org.apache.mahout" per poter calcolare la distanza del coseno.

Il calcolo distribuito interviene generando un istanza di tipo "SparkSession" e le strutture di Apache Spark (rdds, datasets).

#### 4.3.3.1 K-nearest neighbors

La costruzione del KNN ha sfruttato un ciclo che, scorrendo tutto il dataset contenente la base di conoscenza, ovvero le frasi con la relativa categoria, calcola la distanza tra quelle contenute del dataset e quelle da classificare.

Il metodo è stato così implementato:

```
Dataset<Row> datasetMinDistance = cart.map(row -> RowFactory.create(row.getAs("label"),
                    CosineDistanceMeasure.distance(((Vector)row.getAs("phrase")).toArray(),
                    ((Vector) row.getAs("features")).toArray())),
                    RowEncoder.apply(
                    DataTypes.createStructType( new StructField[]{
            DataTypes.createStructField(    "label",
                                            DataTypes.DoubleType,
                                            true),
                                            DataTypes.createStructField("Cos Dist",
                                            DataTypes.DoubleType,
                                            true)})))
                    .groupBy(functions.col("label")).min("CosDist").orderBy("label");

double minDist = datasetMinDistance.agg(functions.min(functions.col("min(CosDist)"))).first().getDouble(0);
```

#### 4.3.3.2 Metodo Ensemble

La costruzione del metodo Ensemble non ha sollevato criticità particolari infatti è bastato effettuare il calcolo della media tra le *accuracy* dei vari algoritmi come in Eq:4.1 con le strutture messe a disposizione da Spark.

$$a_e = \frac{a_{nb} + a_{knn} + a_{svn} + a_{rf} + a_{dnn}}{5}$$

*Eq. 4.1*

# 5. Valutazioni e considerazioni sui risultati

L'addestramento dei modelli viene effettuato utilizzando un dataset di frasi, ciascuna delle quali opportunamente classificata.

Al fine di testare i metodi di classificazione in un ampio studio di tipo *cross validation* sono state selezionate le 10 classi con numerosità più elevata. Nella seguente tabella viene riportato il *tag* (etichetta) e il significato delle 10 classi selezionate:

1. **ATT:** Attivazione SIM, contratti, adsl, etc;
2. **CONFIG:** Problemi di configurazione di telefoni o altri dispositive;
3. **DISATT:** Disattivazioni di linea o modem e recessi da contratto;
4. **FDT:** Informazioni generali: consumo, traffico residuo, etc;
5. **GC:** Dettaglio traffico, esenzioni, agevolazioni;
6. **OFF:** Informazioni su offerte e promozioni;
7. **RIC:** Ricarica;
8. **SERV:** Attivazioni e disattivazioni servizi specifici;
9. **SERVIZIO_CLIENTI:** Richieste di assistenza con operatori;
10. **TS:** Problemi di varia natura (password, ricariche, internet, etc).

Ne conseguono le specifiche di ogni algoritmo di classificazione con le relative accuracy di alcuni casi d'uso nel dominio Telco:

## 5.1 Come attivo offerta

|              | ATT   | CONFIG | DISATT | FDT   | GC    | OFF   | RIC   | SERV  | SERVIZIO_CLIENTI | TS    |
|---:|---|---|---|---|---|---|---|---|---|---|
| **NaiveBayes**   | 0.002 | 0.001  | 0.002  | 0.084 | 0.002 | 0.368 | 0.001 | 0.001 | 0.520            | 0.002 |
| **RandomForest** | 0.000 | 0.000  | 0.000  | 0.000 | 0.000 | 1.000 | 0.000 | 0.000 | 0.000            | 0.000 |
| **DNN**          | 0.000 | 0.000  | 0.000  | 0.000 | 0.000 | 0.990 | 0.000 | 0.000 | 0.009            | 0.000 |
| **SVM**          | 0.000 | 0.000  | 0.000  | 0.000 | 0.000 | 1.000 | 0.000 | 0.000 | 0.000            | 0.000 |
| **KNN**          | 0.000 | 0.000  | 0.000  | 0.000 | 0.000 | 0.000 | 0.000 | 0.000 | 1.000            | 0.000 |
| **ENSEMBLE**     | 0.000 | 0.000  | 0.000  | 0.017 | 0.000 | 0.675 | 0.000 | 0.000 | 0.306            | 0.000 |

*Tabella 5.1 - Accuracy "Come attivo offerta"*

## 5.2 Come configuro la mia offerta

|              | ATT   | CONFIG | DISATT | FDT   | GC    | OFF   | RIC   | SERV  | SERVIZIO_CLIENTI | TS    |
|---:|---|---|---|---|---|---|---|---|---|---|
| **NaiveBayes**   | 0.005 | 0.345  | 0.003  | 0.080 | 0.005 | 0.284 | 0.003 | 0.002 | 0.274            | 0.001 |
| **RandomForest** | 0.000 | 1.000  | 0.000  | 0.000 | 0.000 | 0.000 | 0.000 | 0.000 | 0.000            | 0.000 |
| **DNN**          | 0.000 | 0.994  | 0.000  | 0.000 | 0.000 | 0.005 | 0.000 | 0.000 | 0.000            | 0.000 |
| **SVM**          | 0.000 | 1.000  | 0.000  | 0.000 | 0.000 | 0.000 | 0.000 | 0.000 | 0.000            | 0.000 |
| **KNN**          | 0.000 | 1.000  | 0.000  | 0.000 | 0.000 | 0.000 | 0.000 | 0.000 | 1.000            | 0.000 |
| **ENSEMBLE**     | 0.001 | 0.868  | 0.001  | 0.016 | 0.001 | 0.058 | 0.001 | 0.000 | 0.055            | 0.000 |

*Tabella 5.2 – Accuracy "Come configuro la mia offerta"*

## 5.3 Come ricarico il mio cellulare per riattivare l'offerta

| | ATT | CONFIG | DISATT | FDT | GC | OFF | RIC | SERV | SERVIZIO_CLIENTI | TS |
|---:|---|---|---|---|---|---|---|---|---|---|
| **NaiveBayes** | 0.120 | 0.005 | 0.001 | 0.026 | 0.003 | 0.341 | 0.476 | 0.023 | 0.006 | 0.002 |
| **RandomForest** | 0.000 | 0.000 | 0.000 | 0.000 | 0.000 | 0.000 | 1.000 | 0.000 | 0.000 | 0.000 |
| **DNN** | 0.000 | 0.000 | 0.000 | 0.000 | 0.000 | 0.016 | 0.983 | 0.000 | 0.000 | 0.000 |
| **SVM** | 0.000 | 0.000 | 0.000 | 0.000 | 0.000 | 0.000 | 1.000 | 0.000 | 0.000 | 0.000 |
| **KNN** | 0.000 | 0.000 | 0.000 | 0.000 | 0.000 | 0.000 | 1.000 | 0.000 | 0.000 | 0.000 |
| **ENSEMBLE** | 0.024 | 0.001 | 0.000 | 0.005 | 0.001 | 0.071 | 0.892 | 0.005 | 0.001 | 0.000 |

*Tabella 5.3 - Accuracy "Come ricarico il mio cellulare per riattivare l'offerta"*

Come si evince dagli esempi precedenti i risultati ottenuti dalla somma delle accuratezze espresse dai vari algoritmi (*Ensemble*) garantisce una precisione abbastanza elevata in termini di classificazione della categoria anche su frasi particolarmente entropiche come nell'ultimo esempio.

# 6. Conclusioni

Si evince dai risultati ottenuti che l'applicazione del sistema distribuito in ambito di classificazione è attuabile garantendo risultati più che soddisfacenti in termini di accuratezza dei modelli. Nello specifico l'Ensemble dei risultati ottenuti ha portato percentuali di successo paritari o superiori a quelli riscontrati nell'utilizzo di un singolo elaboratore.

La classificazione, nel suo complesso, ha misurato una componente erratica solo del 1,8% rispetto ai casi presi in esame durante i test pratici.

Diversamente dai precedenti studi su scenari di questa tipologia, grazie all'integrazione di elaborazioni distribuite su più macchine, sono stati superati i limiti riscontrati precedentemente dal punto di vista di costo computazionale.

L'addestramento dei modelli, ora, potrà essere effettuato su dataset di frasi estremamente ampi garantendo, in ogni caso, un'elevata velocità di elaborazione con conseguente riduzione delle tempistiche.

Gli esempi precedenti hanno dimostrato che il dataset composto da un numero estremamente elevato di frasi garantisce, da una parte, maggiore accuratezza in fase di addestramento, e dall'altra, notevoli miglioramenti in fase di predizione della categoria.

# 7. Bibliografia